# A Transportation Digital-Twin Approach for Adaptive Traffic Control Systems

**Sagar Dasgupta***
Ph.D. Student
Department of Civil, Construction & Environmental Engineering
The University of Alabama
3014 Cyber Hall, Box 870205, 248 Kirkbride Lane, Tuscaloosa, AL 35487
Tel: (864) 624-6210; Email: sdasgupta@crimson.ua.edu

**Mizanur Rahman, Ph.D.**
Assistant Professor
Department of Civil, Construction & Environmental Engineering
The University of Alabama
3015 Cyber Hall, Box 870205, 248 Kirkbride Lane, Tuscaloosa, AL 35487
Tuscaloosa, AL 35487
Tel: (205) 348-1717; Email: mizan.rahman@ua.edu

**Abhay D. Lidbe, Ph.D.**
Associate Research Engineer
Alabama Transportation Institute
The University of Alabama
3051 Cyber Hall. Box 870288, 248 Kirkbride Lane, Tuscaloosa, AL 35487
Tel: +1 (205) 393-4031; Email: adlidbe@ua.edu

**Weike Lu, Ph.D**.
Assistant Professor
School of Rail Transportation
Soochow University
No. 8 Jixue Rd., Soochow, Jiangsu, China, 215131
E-mail: weikeguojing@gmail.com

**Steven Jones, Ph.D.**
James R. Cudworth Professor
Department of Civil, Construction & Environmental Engineering
The University of Alabama
3024 Cyber Hall, Box 870205, 248 Kirkbride Lane, Tuscaloosa, AL 35487
Tel: 205-348-3137; Email: sjones@eng.ua.edu

*Corresponding author

**ABSTRACT**
A transportation digital twin represents a digital version of a transportation physical object or process, such as a traffic signal controller, and thereby a two-way real-time data exchange between the physical twin and digital twin. This paper introduces a digital twin approach for adaptive traffic signal control (ATSC) to improve a traveler's driving experience by reducing and redistributing waiting time at an intersection. While an ATSC combined with a connected vehicle concept can reduce waiting time at an intersection and improve travel time in a signalized corridor, it is nearly impossible to reduce traffic delay for congested traffic conditions. To remedy this defect of the traditional ATCS with connected vehicle data, we have developed a digital twin-based ATSC (DT-based ATSC) that considers the waiting time of approaching vehicles towards a subject intersection along with the waiting time of those vehicles at the immediate upstream intersection. We conducted a case study using a microscopic traffic simulation, Simulation of Urban Mobility (SUMO), by developing a digital replica of a roadway network with signalized intersections in an urban setting where vehicle and traffic signal data were collected in real-time. Our analyses reveal that the DT-based ATSC outperforms the connected vehicle-based baseline ATSC in terms of average cumulative waiting time, distribution of drivers' waiting time, and level of services for each approach for different traffic demands and therefore demonstrates our method's superior efficacy.

**Keywords:** Adaptive signal control systems, digital twin, signalized intersection, connected vehicles, and waiting time

## INTRODUCTION

Traffic congestion is a persistent challenge to urban transportation systems and for the transportation professionals responsible for their operation. Literature suggests that adaptive traffic signal control (ATSC) systems have the potential to improve signalized intersectional performance by reducing delays, travel times, and queues and thus generating additional capacity and improving the overall intersectional level-of-service (LOS) (*1*, *2*). However, such performance improvements are limited by the traffic data that is collected using inductive loops and other conventional sensor systems (*3*). Real-time signal optimization at signalized intersections is poised to undergo substantial changes with the availability of trajectory data from connected vehicle (CV) technologies. CV technology, also referred to as vehicle-to-vehicle (V2V) and vehicle-to-infrastructure (V2I) communication, can be defined as the rapid and continuous electronic communication between vehicles and infrastructure that readily improves safety and mobility and reduces the environmental impact of transportation (*4*).

As industry slowly shifts towards intelligent or smart transportation systems, digital twin technology can further revolutionize traffic management and operations. A digital twin (DT) is a digital version of a physical object or process based on two-way data exchange between digital and physical entities in real-time. The concept of creating "twins" is to help improve decision-making. The transportation digital twin can be conceptualized as traffic data being collected from different physical systems, such as sensors, connected vehicles (CV), traffic signals, and traffic monitoring cameras in real time to create a cyber copy of the systems. Although the concept of DT replicates the idea of cyber-physical systems (CPS), transportation DTs are expected to leverage the embedded sensor systems of physical transportation systems to provide real-time and time sensitive transportation services instead of focusing only on the applications of the CPS domain. The primary challenge to achieve this is combining and linking data from heterogeneous sources of the physical systems to create a cyber copy of the real-world traffic operations for real-time traffic management.

DT technology is still conceptually evolving and has no single standard definition. For example, Steyn and Broekman (*5*) define DT as integrated multi-physics, multiscale, and probabilistic simulations of a complex product in manufacturing industries that mirrors the behavior and environmental responses of its corresponding physical twin. Rudskoy et al. (*6*) understand DT as a module that reproduces a detailed digital model of the road and allows for modeling and experiments to test solutions and simulate different situations. DT has been increasingly adopted in several fields, such as manufacturing and production engineering (*7*, *8*), medicine (*9*), healthcare (*10*), systems engineering (*11*), and product design (*12*). However, DT applications in transportation engineering has thus far been limited and is considered to be in its nascent stages – and its application towards traffic signal control problems is currently non-existent. The objective of this paper is to develop a DT-based ATSC strategy for improving driving experience in an urban roadway network and therefore set a new line of research going forward. As such, this paper presents a proof-of-concept application of DT to ATSC operations at a signalized intersection.

## RELATED WORK

ATSCs have evolved from basic adaptive features (e.g., timing modifications) to more real-time adaptation as the underlying algorithms and sensor technologies have improved. For example, early ATSC (which were mainly proprietary) used basic traffic data from the inductive loops and/or traffic cameras to gradually adapt to changing traffic conditions and generally improve operational performance of signalized intersections (*1–5*). Other ATSCs used more advanced algorithms to improve effectiveness (*6–8*). Advancements in machine learning techniques and wireless communication technologies have further improved ATSC capabilities in responding to dynamic traffic conditions in near real-time (*9–13*). Yet, the most advanced ATSC strategies were often not feasible due to their operational complexity and high-resolution detection requirements (*14*). CV and/or V2I technologies are helping fill this gap. Thus, more recent ATSCs use vehicle trajectory data from CV/V2I environments to refine the ATSC performance further. For example, a new predictive traffic control algorithm that employs robust trajectory data was developed using a rolling-horizon strategy in which the phasing was chosen to optimize an objective function for a 15-s period in the future (*14*). Vehicular ad hoc networks were used to collect and aggregate real-time speed and position information on individual vehicles to optimize signal control at traffic intersections (*15*). The optimization problem was formulated as a job scheduling problem on processors, with discrete jobs corresponding to individual vehicle platoons. Another study developed a delay and weighted-delay-based algorithm to optimize signal timings at an isolated intersection (*16*). Aljaafreh and Al-Oudat (*17*) developed a novel multi-agent-based control method for an integrated network of ATSCs within a V2I communication environment. Bhave et al. (*18*) presented a real-time adaptive signal phase allocation algorithm using connected vehicle data that optimizes the phase sequence and duration by solving a two-level optimization problem.

Future transportation is believed to be greatly benefited by the addition of real-time information gathering and analytics, along with the proliferation of machine and deep learning algorithms to the existing intelligent transport systems (ITS) (*19*). Using DT within the correct framework creates the potential to anticipate possible traffic problems (*20*). In fact, the practical feasibility of modeling road traffic using a DT has been verified (*21*), and its potential to enable AVs as a crucial component of smart mobility has also been documented (*22*). While the DT concept is still evolving, Aslani et al. (*23*) developed a DT simulation model capable of providing traffic performance measures in near real-time. The study also revealed the data unavailability necessary to leverage real-time high frequency connected corridor data streams for real-time applications. Yet another a study demonstrated the feasibility of DT-assisted real-time traffic data prediction methods for analyzing traffic flow and velocity data monitored by sensors and transmitted via 5G (*24*). Goodall et al. demonstrated the successful implementation of DT on a local road that accurately measured road characteristics and provided data to assist in road (*25*). Finally, the principles of developing DT for producing active vehicle safety systems exemplified by the braking system has also been studied (*26*).

While the above literature suggests that DT has the potential to offer transformative capabilities, the current body of literature lacks sufficient evidence for the application of DT towards direct traffic signal control. Thus, this study presents a proof of concept of the application of DT technology to allocate signal phase and timing of ATSC. The primary contribution of this paper is to develop a DT-based ATSC strategy that improves the user experience in an urban roadway network by reducing delays in the network. Unlike the trajectory data, whose availability to ATSC is limited to certain spatial limits from the intersection, the DT setup relaxes such limits which makes the tracking of vehicle delays beyond a single intersection possible. The DT-based ATSC does not address the vehicle routing optimization problem. Instead, it prioritizes a green signal for the approach with vehicles experiencing the highest accumulated waiting time (AWT). The AWT for each vehicle is calculated as the sum of all the delays a vehicle accrues while navigating the network before leaving the subject intersection. This includes delays before arriving and delays accrued at the intersection. Thus, with the DT-based ATSC strategy, vehicles with longer AWT (i.e., poor user experience) receive priority at the subject intersection. The novelty of the proposed approach is that it can distribute waiting time throughout a signalized network to provide a better user experience in congested traffic conditions, and is scalable for a city-wide network. In this paper, however, the authors demonstrate the proof-of-concept for a DT-based ATSC strategy for a single signalized intersection.

## DIGITAL TWIN BASED ADAPTIVE TRAFFIC SIGNAL CONTROL FRAMEWORK

This paper introduces a novel DT approach for an ATSC to improve user experiences by reducing waiting time at an intersection. DT of a traffic signal controller can be defined as a cyber copy of a physical system, in this case an ATSC, representing a complete set of functionalities of for real traffic signal controller. **Figure 1** presents a high-level view of the proposed digital and physical twins of ATSC. In this approach, it is necessary to create a DT of a traffic signal controller along with vehicles and road networks. Vehicle trajectories and signal data (e.g., phasing and timing) must be transmitted from a physical twin to the corresponding DT through wired and wireless communications. After obtaining the data, the proposed DT-based algorithm generates traffic signal phase and timing parameters, which are then relayed to the real-world traffic signal controller to change signal heads accordingly. The computation module of the DT processes the collected vehicle trajectory and traffic signal data to notify the physical systems about the findings or sends control commands to make necessary changes to the signal heads. Unlike traditional ATSC with CV data, a DT system can mirror the city-wide roadway network, vehicles, and traffic signals instead of representing CVs only at isolated signals.

Although an ATSC combined with a CV concept can reduce waiting time at an intersection *(27)* and improve travel timing significantly in a signalized corridor, it is nearly impossible to reduce traffic delay for saturated traffic conditions. Unlike traditional ATSC with CV data, the proposed method utilizes both the cumulative waiting time of approaching vehicles towards an intersection and the AWT of these vehicles at the immediate upstream intersection, as shown in **Figure 2(a)**. At any intersection, the green time is prioritized based on the average of vehicle AWT for a given movement. Thus, the DT-based strategy prioritizes and allocates green time for a particular phase depending on the respective user experience in terms of cumulative AWT at both the subject intersection and immediate upstream intersection. This approach can distribute waiting time throughout the signalized network thus providing a better travel experience in congested traffic. This concept can be expanded and scaled up to a city-wide network, as shown in **Figure 2(b)**.

Digital Twin

Digital replica of vehicles, traffic signals and roadway networks

Collecting vehicles and traffic signal phase and timing data

Adaptive traffic signal control algorithm

Generation of traffic signal phase and timing information

Decision Making for traffic signal controller

Wired/Wireless Communication

Sensing of vehicles' trajectory data and traffic signal timing and phasing information

Actuation information of traffic signal controller

Physical replica or real vehicles, traffic signals and roadway networks

Physical Twin

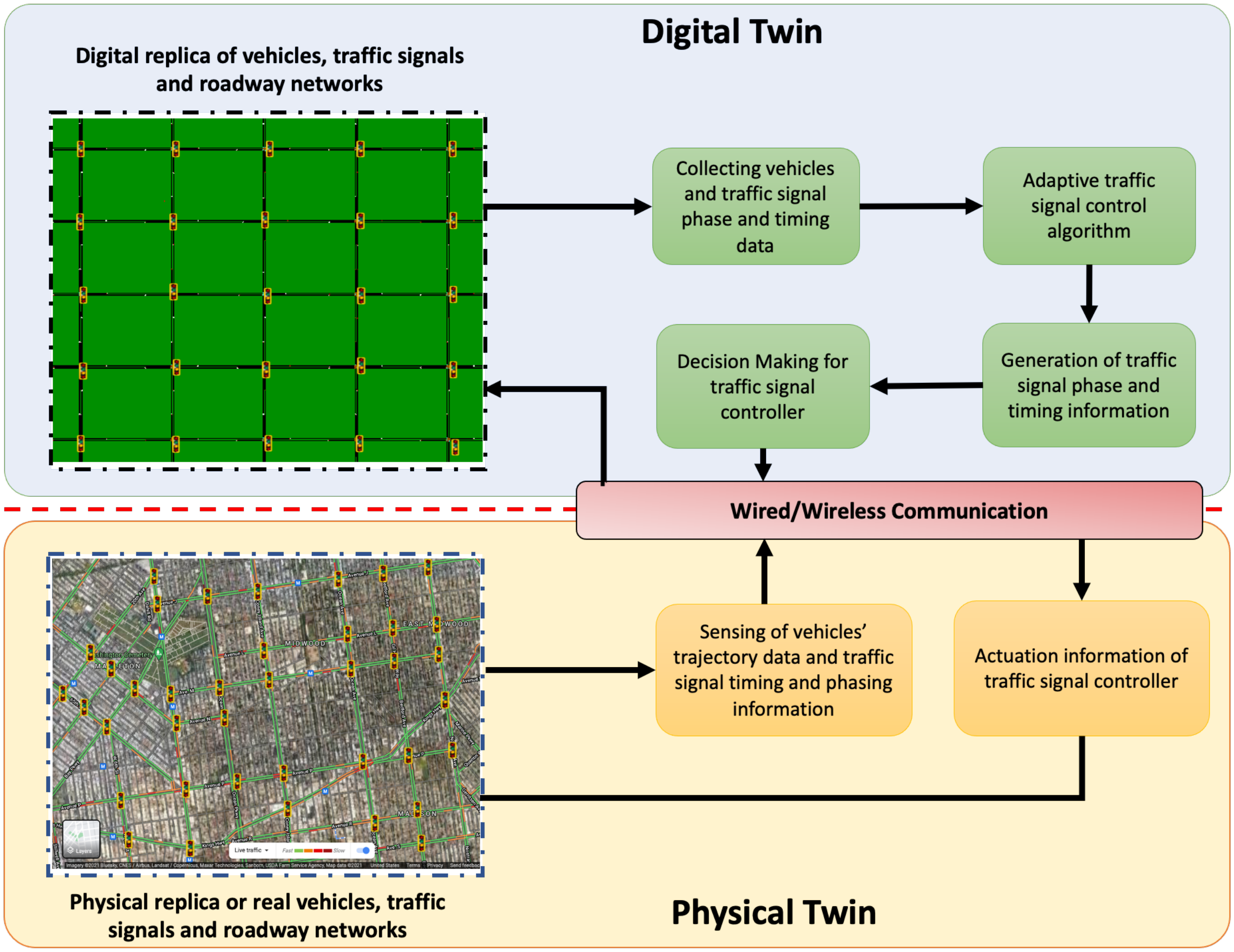

**Figure 1 Twining of digital and physical space of adaptive traffic signal control systems**

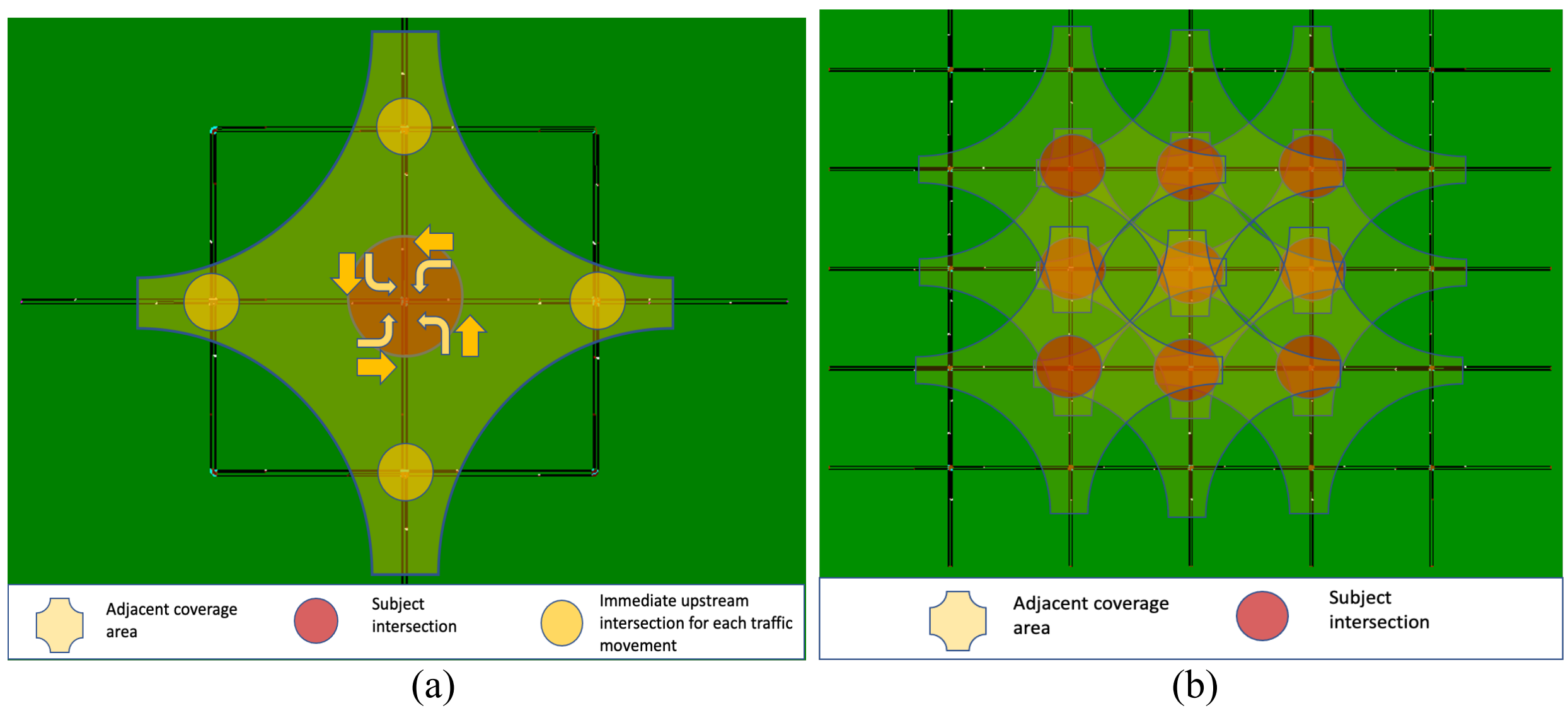

(a) (b)

**Figure 2 Digital twin conceptual architecture for improving user (driving) experience in a high-density urban roadway network: (a) for a signalized intersection; (b) city-wide signalized intersections**

As shown in **Figure 3**, four vehicles (VID_1, VID_2, VID_3 and VID_4) are approaching the subject intersection from an immediate upstream intersection. VID 1 and VID 2 are the through traffic from the immediate upstream intersection, while VID_3 and VID_4 are the right- and left-turning traffic, respectively. When these four vehicles reach the subject intersection, the sum of all waiting time faced while traveling from the immediate upstream intersection to the subject intersection (i.e., AWT) is considered. The AWT of the vehicles on their arrival at the subject intersection are defined as $AWT_1$, $AWT_2$, $AWT_3$ and $AWT_4$. The vehicle is considered to incur wait time if its speed is 0.1 m/s or lower (as per the SUMO user guide *(37)*). Equation 1 is used to calculate AWT waiting time for each vehicle.

$$AWT_t = \sum_{i=t-t'}^{t} WT_i\,; \qquad \forall\, i = (t-t',\ t-(t'+1), t-(t'+2), \ldots\ t) \tag{1}$$

where $t'$ is the time when a vehicle starts approaching the immediate upstream intersection, $t$ is the time when a vehicle starts departing the subject intersection, $WT_i$ represents the waiting time at time $i$, and $AWT_t$ is the AWT at time $t$.

The average accumulated waiting time ($AAWT$) of $n$ vehicles is derived using the following equation 2.

$$AAWT_t = \frac{\sum_{j=1}^{n}(AWT_1 + AWT_2 + AWT_3 + \cdots + AWT_n)}{n}\ ;\ \forall\, j = 1, 2, 3, \ldots \ldots n \tag{2}$$

where $AAWT$ represents the average delays in seconds/vehicle, $AWT$ is the vehicle delay in seconds, and $n$ is the total number of vehicles that are approaching towards a subject intersection. $n$ represents the number of vehicles on respective approaches for calculating $AAWT$ for each approach and represents the total number of vehicles approaching the subject intersection while calculating the intersectional $AAWT$. The DT-based algorithm allocates green to a movement (through or left-turns) with the maximum $AAWT$ at time $t$ compared to other movements. Thus, the phase for green allocation is decided as:

$$max(AAWT_t^M);\ where, M = EBL, EBT, WBL\ WBT, NBL, NBT, SBL, and\ SBT$$

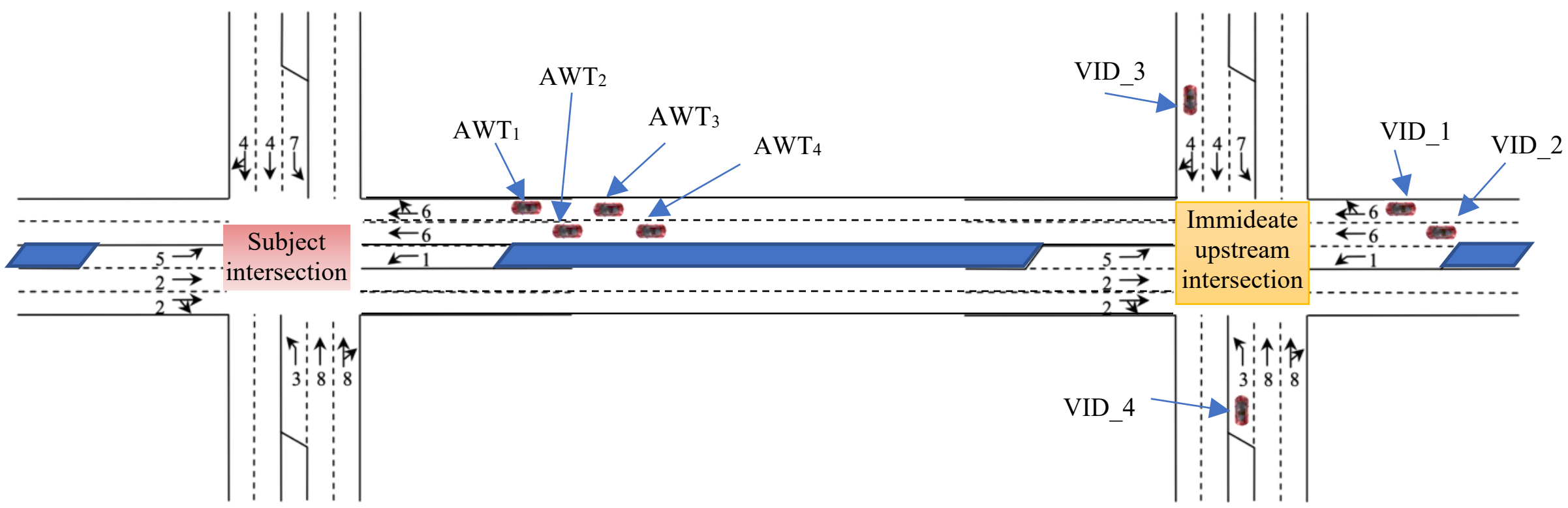

**Figure 3 An example scenario on how to calculate the average waiting time (AWT)**

The goals of the DT-based ATSC are as follows: (i) improve the user experience by reducing waiting time at the subject signalized intersection; (ii) respond to real-time traffic demands and allocate traffic signal phase by considering waiting times at the immediate upstream intersections; and (iii) scale to a city-wide signalized roadway network. To achieve these goals, a rolling-horizon strategy has been developed wherein the signal controller allocates green phase for an approach with vehicles with the highest AWT. **Figure 4** presents an overview of the DT-based ATSC logic. The algorithm runs in real-time evaluating and recording the wait time and AWT of each vehicle at the subject and the immediate upstream intersections every 1 second. The phase change decision is evaluated after every 5 seconds from the start of the green phase. A phase change, when warranted, follows through a yellow change interval of 2 seconds and within 1 second of all-red clearance. A phase change is not warranted if the approach currently has green and still has maximum AWT after 5 seconds from the start of the last green. The yellow of 2 seconds and red of 1 second is skipped in such cases.

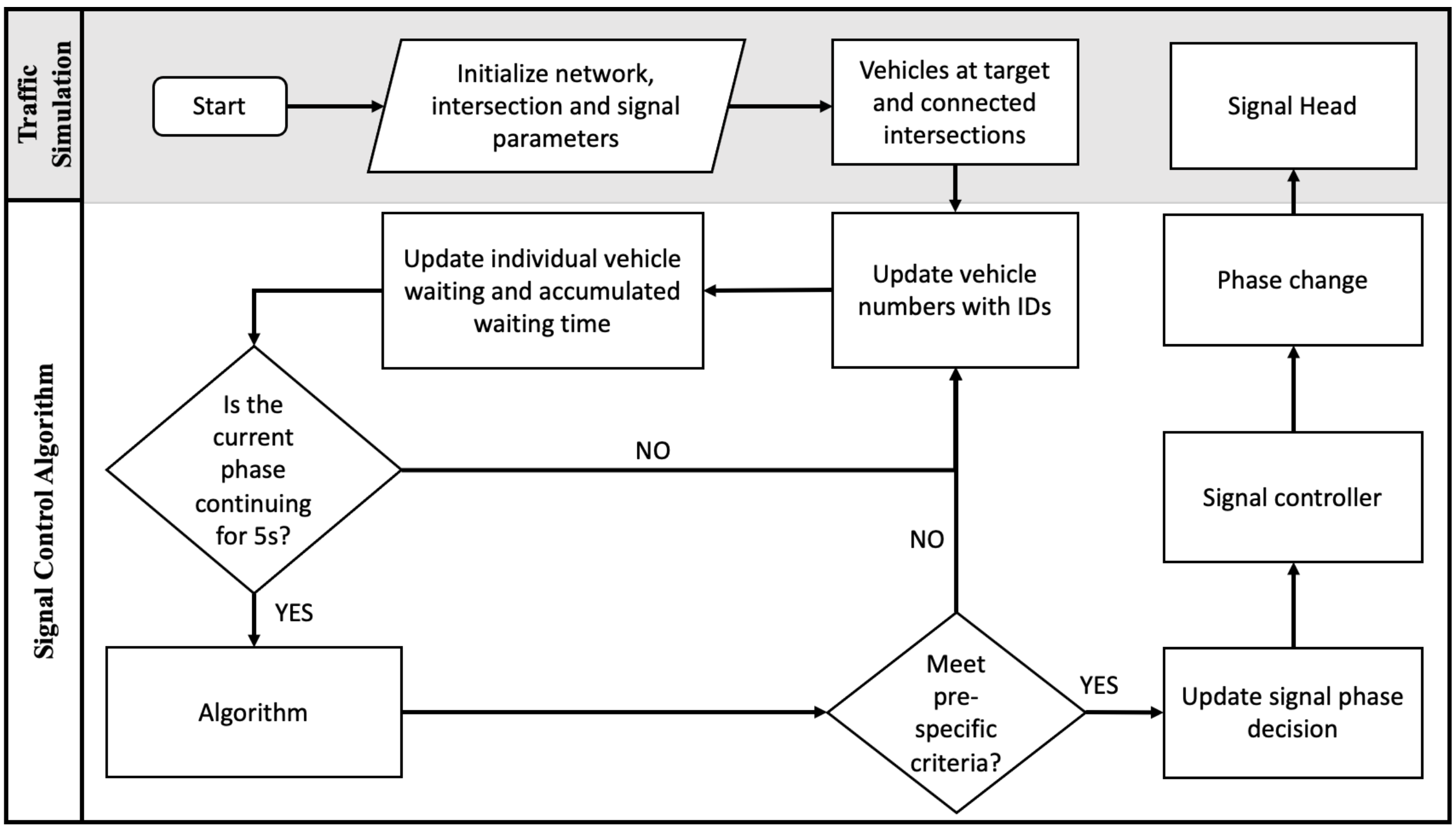

**Figure 4 Overview of digital-twin based adaptive traffic signal control logic**

## A CASE STUDY

To demonstrate the efficacy of our DT-based ATSC approach, this section presents a case study using a microscopic traffic simulator, Simulation of Urban Mobility (SUMO) *(37,* to mimic traffic in a typical urban signalized network. The traffic signal control algorithm runs through its Traffic Control Interface (TraCI) interface *(38)*.

### Simulation Set-up

**Figure 5(a)** presents a screenshot from the SUMO graphical user interface showing the geometry of nodes and edges. A node represents an intersection, and an edge refers to a roadway segment connecting the two nodes. Dedicated left-turn pockets are used at the junction for the left-turn movements. The connections which define the possible directions a vehicle can take when it arrives at a junction are shown in **Figure 5(a)**. The connections are the same for all the junctions in the roadway network. **Figure 5(b)** presents the different components of the SUMO simulation environment. NETEDIT module is used to define the vehicular demand in a SUMO engine. Multiple flows (from-to) are created by selecting peripheral start and end edges for each flow. The NETEDIT module by default uses the shortest minimum path between the start and end edge for the flow creation and the path can pass through any edges of the network. Default vehicle types are used for the simulation. The simulation does not include any person or pedestrians. Different traffic demands are created by changing the vehicle generation per hour (vph) in the flow attributes. The traffic demand values used in this series of simulations are 250, 500, 750, and 1000 vph. TraCI is used for collecting vehicle level data and controlling the signal controller of the target intersection. TraCI inbuilt functions were used to collect vehicle ID and associated waiting and accumulated waiting time during the simulation. The collected data is then used inputs to the algorithm, and the traffic signal heads are controlled based on the algorithm output. All simulation runs were conducted for a period of 3,600 seconds, including the 600 seconds of warm-up and another 600 seconds of cool-down. The results reported henceforth relate to the 2,400 seconds of actual simulation period only.

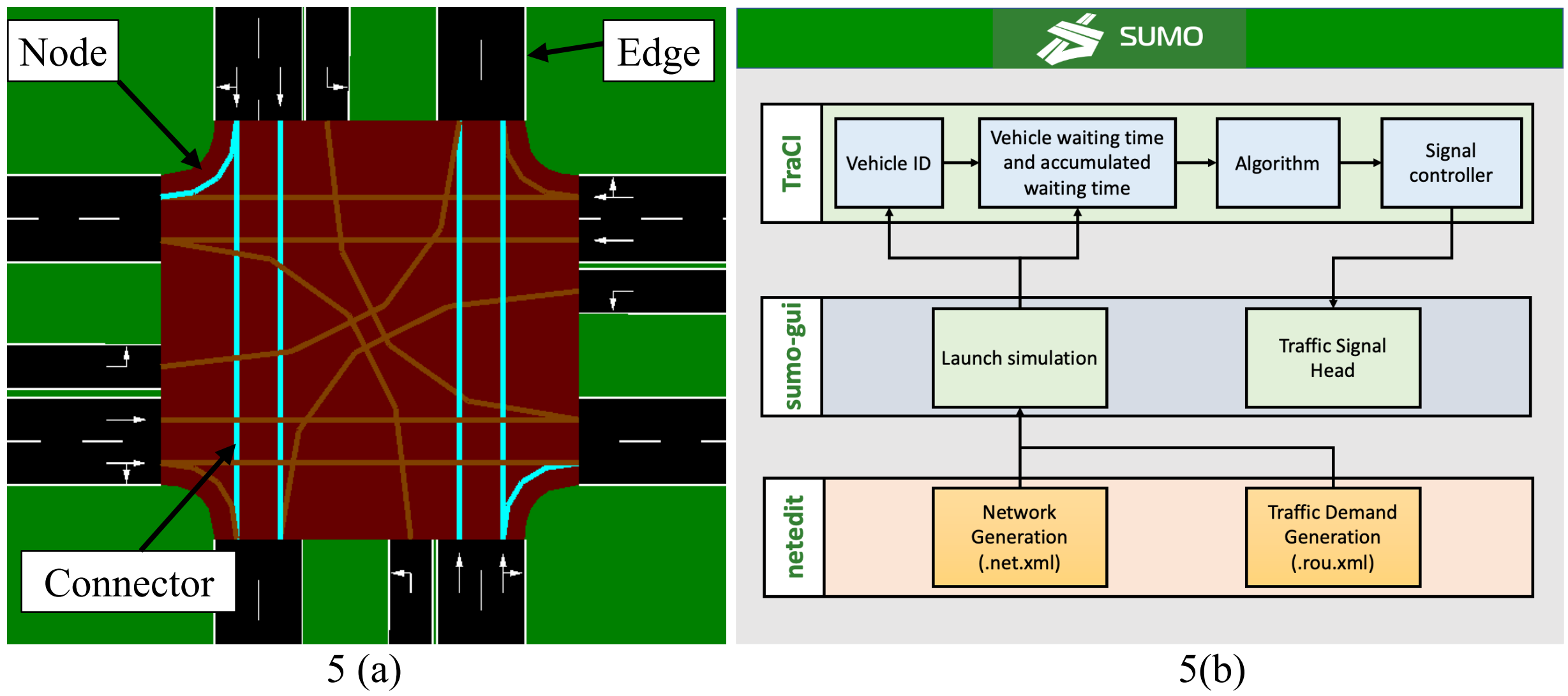

5 (a) 5(b)

**Figure 5 Simulation of digital twin environment using Simulation of Urban Mobility (SUMO) (a) Connections at junctions; and (b) simulation Environment**

## Evaluation Scenarios

To evaluate the performance of the DT-based ATSC, a baseline scenario was used for comparison purposes. **Table 1** presents different concepts, traffic movements, and traffic demands specified for the two baseline and test scenario. The CV-based ATSC baseline scenario assumes that trajectory data only for the subject intersection is available to the signal controller. Hence, the AWT in the baseline scenario was calculated based on waiting times at the subject intersection only and ignored those at the immediate upstream intersections. Four different traffic demands — 250 vph, 500 vph, 750 vph, and 1000 vph — were selected to simulate the subject intersection at different levels of service (LOS) for evaluating the DT-based ATSC performance.

**TABLE 1 Evaluation Scenarios**

<table>
<tr><th>Scenarios</th><th>Concept</th><th>Traffic Movements</th><th>Traffic Demand</th></tr>
<tr><td><b>Baseline ATSC</b></td><td>All the approaching vehicles towards a subject intersection could exchange vehicle trajectory data with the controllers through wireless connectivity</td><td rowspan="2">• East Bound Through (EBT)<br>• West Bound Through (EBT)<br>• South Bound Through (EBT)<br>• North Bound Through (EBT)<br>• East Bound Left (EBL)<br>• West Bound Left (EBL)<br>• South Bound Left (EBL)<br>• North Bound Left (EBL)</td><td rowspan="2">The following are the traffic demand at each entry point of the road network:<br>• 250 vph<br>• 500 vph<br>• 750 vph<br>• 1000 vph</td></tr>
<tr><td><b>Digital Twin-ATSC</b></td><td>All the vehicles in the network and traffic signal controllers can exchange vehicle trajectory data and signal phase and timing information with a backend computing infrastructure (e.g., cloud infrastructure or roadside data infrastructure)</td></tr>
</table>

**Evaluation Metrics**

The study uses AWT and LOS for evaluating the performances of the two signal control regimes. For the former, both distribution and cumulative distributions of the same are analyzed to evaluate if there are significant differences with the DT-based approach. The comparison between the skewness of the AWT distributions indicates the superiority of a particular signal control strategy over the other. The cumulative distribution indicates the change in the AWT as well as the total AWT accrued over the simulation time period. The LOS indicates the overall condition of each approach and for the whole intersection.

**Evaluation Outcome**

**Figure 6** presents a comparison between baseline and DT scenarios for through and left-turn movements in terms of cumulative waiting time (the results for only 750 vph scenario are presented here). As shown in **Figure 6(a)**, the difference of cumulative waiting for the SBT, NBT, and EBT approach between baseline and DT scenarios significantly increases over time, and the baseline scenario shows a much higher cumulative waiting time as the simulation time increases. However, for the WB approach, the cumulative waiting time for the DT-based ATSC scenario slowly increases over the simulation period. A possible reason for this might be low traffic in the WBT approach over the simulation period. Our DT-based ATSC always tries to allocate green time depending on the maximum average waiting time between all the traffic movements. On the other hand, we found that the difference of cumulative waiting between baseline and DT-based ATSC scenarios markedly increases over time for all the left-turn movements, and the baseline approach shows a much higher cumulative waiting time when increasing the simulation time. This indicates lower waiting time at the subject intersection and proves the efficacy of our DT-based ATSC scenario.

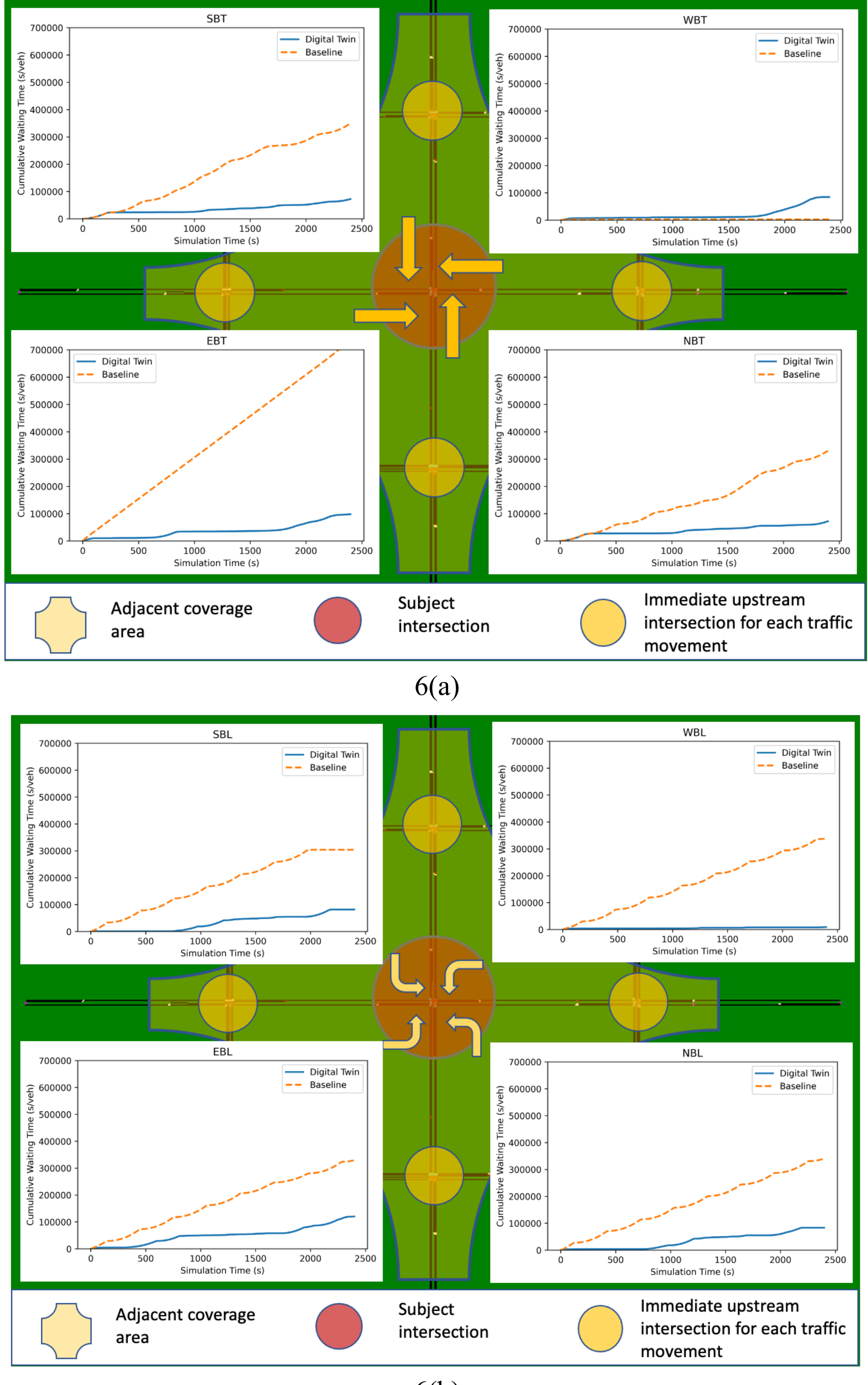

6(a)

6(b)

**Figure 6 Cumulative AWT (750 vph): (a) Through movements (b) Left-turn movements**

**Figure 7** shows the frequency distribution of AWT for the baseline and DT-based ATSC scenarios for through and left turn movements (for 750 vph). As shown in **Figure 7(a)**, the distribution of vehicle waiting for SBT and NBT is more skewed towards lower waiting times for the DT-based ATSC scenario compared to the baseline scenario. Although EBT and WBT approaches show a similar pattern for the distribution, the baseline scenario shows a higher frequency with a lower waiting time. However, the waiting time distribution for the baseline scenario does not follow a decreasing pattern, and instead exhibits a random pattern. The mean and standard deviation of AWT is much less for SBT and NBT using the DT-based ATSC scenario compared to the baseline scenario. On the other hand, EBT and WBT show higher mean and standard deviation of AWT for the DT-based ATSC scenarios compared to the baseline approach. Similarly, the distribution of AWT for all left turn movements are more skewed towards a lower waiting time for DT-based ATSC scenarios compared to baseline scenarios summarized in **Figure 7(b)**. Also, the mean and standard deviation are lower for the DT-based ATSC scenario compared to the baseline scenario.

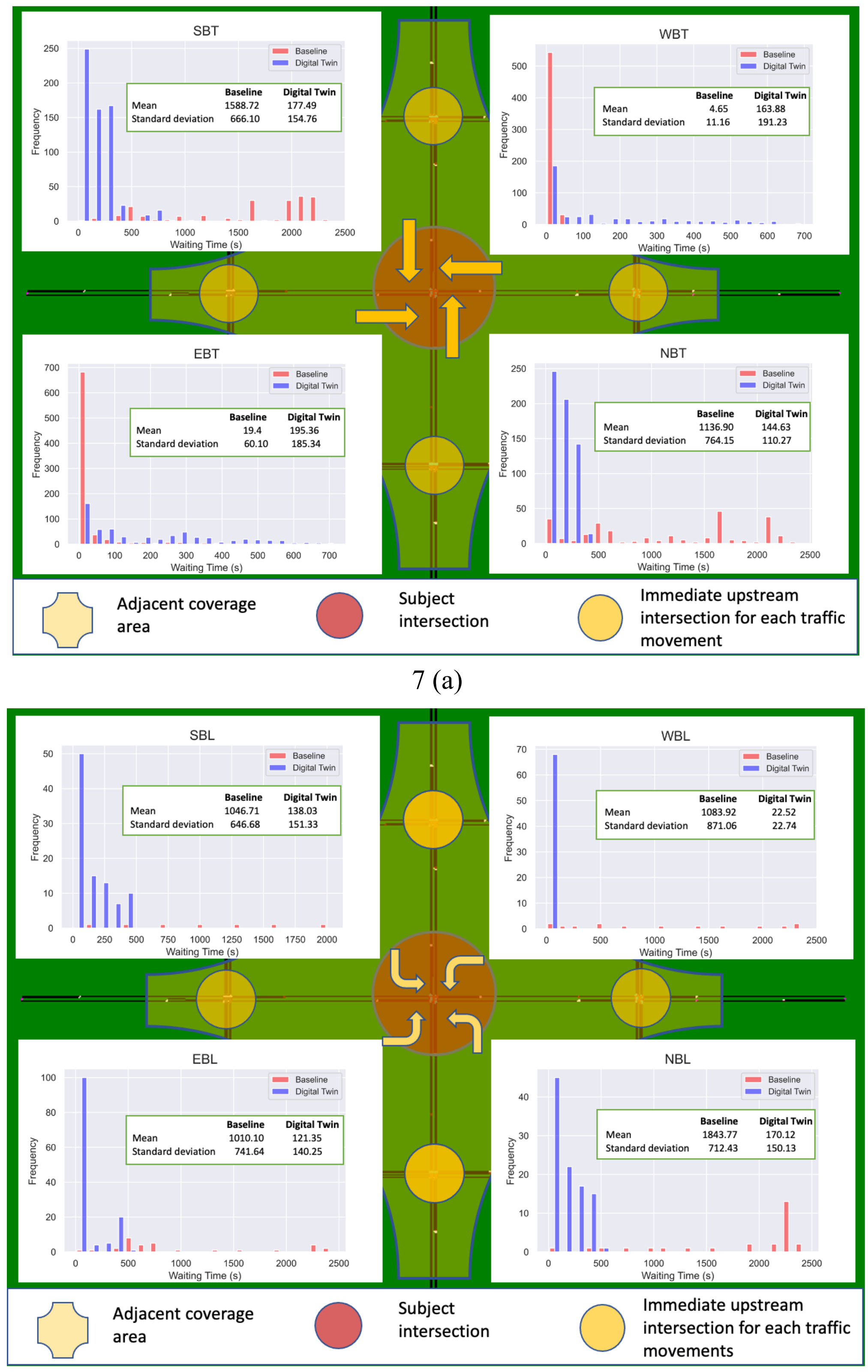

**Figure 7 AWT distribution (750 vph): (a) Through movements (b) Left turn movements**

**Figure 8** shows the frequency distribution of the AWT for all movements combined for the baseline and DT-based approach for the subject intersection. The figure shows that the distribution is more skewed towards lower AWT for the DT-based ATSC scenario compared to the baseline scenario. Unlike the DT-based approach, the AWT distribution for the baseline approach does not follow a decreasing pattern and displays a random pattern. Furthermore, the mean and standard deviation of AWT is lower with the DT-based ATSC scenario than the baseline approach. Overall, the DT-based ATSC scenario can be thought of as redistributing the AWT among different approaches and consequently providing a better user experience at the subject intersection.

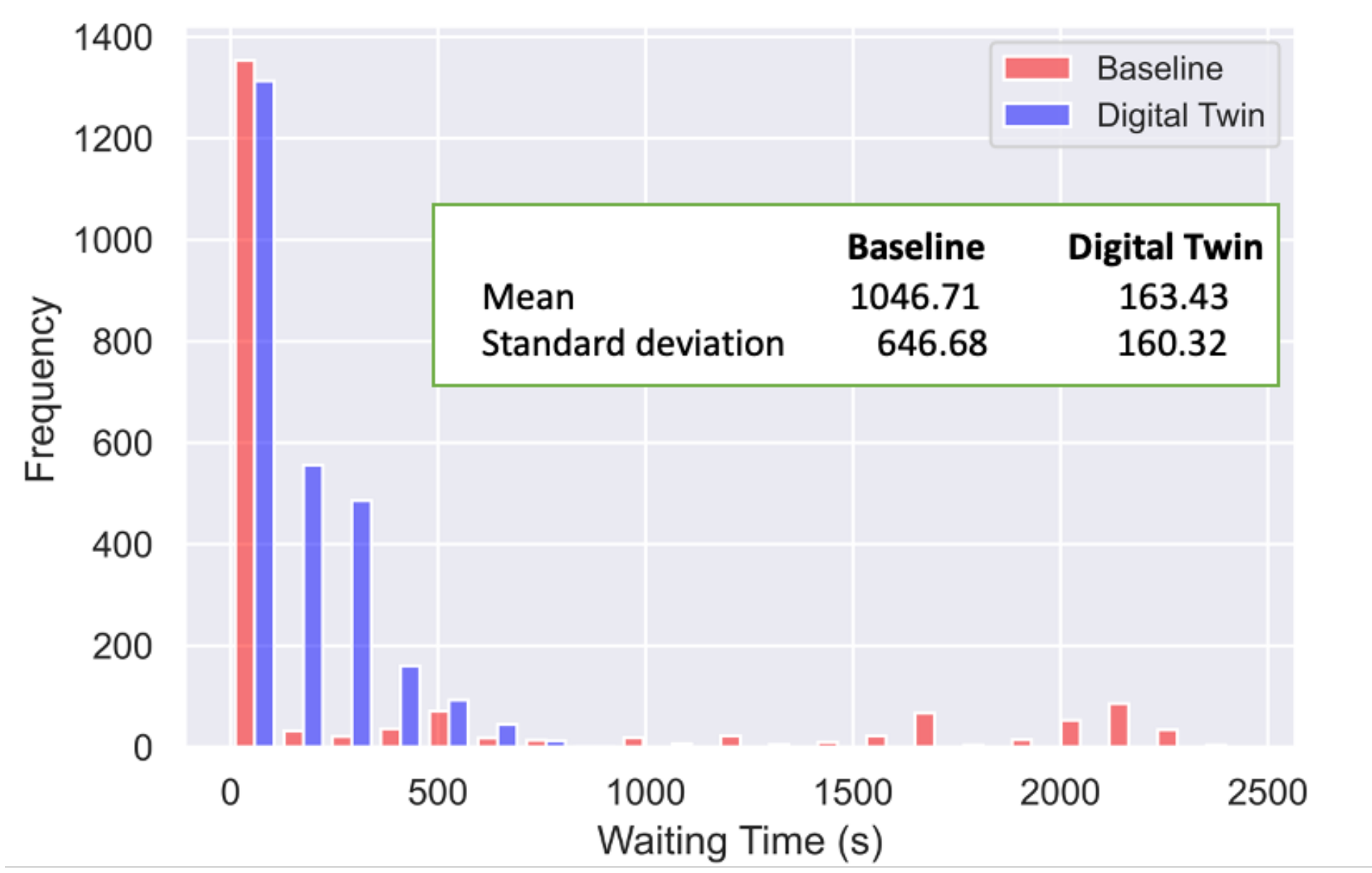

**Figure 8 Comparison between waiting time distribution for the overall intersection (750 vph; Left Turn Traffic Only; 5 seconds green time)**

**Table 2** presents a summary of the average delay per vehicle and corresponding level of service (LOS) *(39)* for each of the subject intersection movements for the four different traffic demands, i.e., 250 vph, 500 vph, 750 vph and 1000 vph, at the entry points of the simulated network. Of the total 32 cases, 28 show better or the same LOS for the DT-based ATSC scenario. Only four cases show lower LOS compared to the baseline scenario. While there are 11 cases where the baseline scenario shows LOS F, there are no such cases for the DT-based ATSC strategy. Overall, it can be concluded that the DT-based ATSC scenario offers superior performance compared to the baseline CV scenario for different traffic demands.

**TABLE 2 Summary of Average Delay and Level of Service (LOS) for Each Approach**

| Average Delay (s/veh) [Level of Service] | | | | | | | | | | |
|---|---|---|---|---|---|---|---|---|---|---|
| **Traffic demand (vph)** | **Scenarios** | **WBL** | **WBT** | **NBL** | **NBT** | **EBL** | **EBT** | **SBL** | **SBT** | **Overall** |
| **250** | Baseline | 51.3 [D] | 21.9 [C] | 22.2 [C] | 54.6 [D] | 36.7 [D] | 90.8 [F] | 10.7 [B] | 54.2 [D] | 42.8 [D] |
| | **Digital Twin** | **21.7 [C]** | **19.0 [B]** | **25.2 [C]** | **30.5 [C]** | **42.3 [D]** | **26.9 [C]** | **3.5 [A]** | **23.5 [C]** | **24.1 [C]** |
| **500** | Baseline | 24.2 [C] | 10.2 [B] | 16.3 [B] | 40.0 [D] | 32.0 [C] | 59.8 [E] | 3.2 [A] | 34.8 [C] | 27.6 [C] |
| | **Digital Twin** | **26.9 [C]** | **18.7 [B]** | **15.8 [B]** | **28.1 [C]** | **46.6 [D]** | **25.0 [C]** | **10.1 [B]** | **34.0 [C]** | **25.7 [C]** |
| **750** | Baseline | 141.2 [F] | 1.2 [A] | 142.0 [F] | 137.5 [F] | 137.3 [F] | 306.5 [F] | 126.7 [F] | 145.2 [F] | 142.2 [F] |
| | **Digital Twin** | **3.9 [A]** | **35.3 [D]** | **34.8 [C]** | **30.2 [C]** | **50.1 [D]** | **40.9 [D]** | **34.1 [C]** | **30.2 [C]** | **32.4 [C]** |
| **1000** | Baseline | 65.0 [E] | 2.9 [A] | 74.2 [E] | 95.4 [F] | 70.0 [E] | 160.8 [F] | 12.9 [B] | 95.2 [F] | 72.1 [E] |
| | **DIGITAL Twin** | **1.7 [A]** | **42.8 [D]** | **36.0 [D]** | **26.6 [C]** | **50.0 [D]** | **55.4 [E]** | **34.7 [C]** | **16.2 [B]** | **32.9 [C]** |

## CONCLUSIONS

In this study, a proof-of-concept of DT-based ATCS was developed to reduce waiting time at an intersection and improve user experience. Unlike traditional ATSC with CV trajectory data, this study utilizes the waiting time of approaching vehicles towards the subject intersection along with the waiting time accrued at the immediate upstream intersection. The AAWT metric is used for prioritizing and allocating green time to an approach. The novelty of the proposed approach is that it can distribute waiting time throughout a signalized network to provide a better travel experience in a congested traffic condition and is scalable for a city-wide network. A case study is conducted using a microscopic traffic simulation in SUMO to develop a digital replica of a signalized network in an urban area where vehicle and traffic signal data were collected in real-time. The analyses reveal that the difference in cumulative AWT for the baseline CV scenario is higher compared to that of DT. In addition, it is observed that the frequency distribution of AWT is more skewed towards lower waiting time for DT-based ATSC as compared to the CV-based baseline approach. In addition, the mean and standard deviation of AWT is lower using the DT-based ATSC compared to the baseline approach and exhibits shorter waiting periods for the users at the subject intersection thus affirming the efficacy of the DT-based ATSC. Moreover, we found that the DT-based ATSC shows better or the same LOS compared to the baseline scenario for different traffic demands. Overall, the DT-based ATSC outperforms the baseline scenario for different traffic demands. In our follow-up study, the scalability of our approach will be investigated in a city-wide signalized roadway network and evaluated the efficacy of our approach.

## ACKNOWLEDGMENTS

This material is based on a study supported by the Alabama Transportation Institute (ATI). Any opinions, findings, and conclusions or recommendations expressed in this material are those of the author(s) and do not necessarily reflect the views of the Alabama Transportation Institute (ATI).

Matthew Shrode Hargis of the University of Alabama provided valuable assistance in editing the paper.

## AUTHOR CONTRIBUTIONS

The authors confirm contribution to the paper as follows: study conception and design: S. Dasgupta, M. Rahman, and A. Lidbe; data collection: S. Dasgupta, M. Rahman, A. Lidbe and W. Lu; interpretation of results: S. Dasgupta, M. Rahman, A. Lidbe, and S. Jones; draft manuscript preparation: S. Dasgupta, M. Rahman, A. Lidbe, W. Lu and S. Jones. All authors reviewed the results and approved the final version of the manuscript.

## REFERENCES

1. Stevanovic, A., N. Dobrota, and N. Mitrovic. Benefits of Adaptive Traffic Control Deployments-A Review of Evaluation Studies. 2019.
2. Stevanovic, A. *NCHRP Synthesis 403 Adaptive Traffic Control Systems : Domestic and Foreign State of Practice*. Washington, D.C., 2010.
3. Wang, Y., X. Yang, H. Liang, and Y. Liu. A Review of the Self-Adaptive Traffic Signal Control System Based on Future Traffic Environment. *Journal of Advanced Transportation*, Vol. 2018, 2018. https://doi.org/10.1155/2018/1096123.
4. U.S. Department of Transportation. *Connected Vehicles: Benefits, Roles, Outcomes*. Washington DC, 2015.
5. Steyn, W. J., and A. Broekman. Process for the Development of a Digital Twin of a Local Road – A Case Study. 2021, pp. 11–22. https://doi.org/10.1007/978-3-030-79638-9_2.
6. Rudskoy, A., I. Ilin, and A. Prokhorov. Digital Twins in the Intelligent Transport Systems. *Transportation Research Procedia*, Vol. 54, 2021, pp. 927–935. https://doi.org/10.1016/J.TRPRO.2021.02.152.
7. Schleich, B., N. Anwer, L. Mathieu, and S. Wartzack. Shaping the Digital Twin for Design and Production Engineering. *CIRP Annals*, Vol. 66, No. 1, 2017, pp. 141–144. https://doi.org/10.1016/J.CIRP.2017.04.040.
8. Kritzinger, W., M. Karner, G. Traar, J. Henjes, and W. Sihn. Digital Twin in Manufacturing: A Categorical Literature Review and Classification. *IFAC-PapersOnLine*, Vol. 51, No. 11, 2018, pp. 1016–1022. https://doi.org/10.1016/J.IFACOL.2018.08.474.
9. Corral-Acero, J., F. Margara, M. Marciniak, C. Rodero, F. Loncaric, Y. Feng, A. Gilbert, J. F. Fernandes, H. A. Bukhari, A. Wajdan, M. V. Martinez, M. S. Santos, M. Shamohammdi, H. Luo, P. Westphal, P. Leeson, P. DiAchille, V. Gurev, M. Mayr, L. Geris, P. Pathmanathan, T. Morrison, R. Cornelussen, F. Prinzen, T. Delhaas, A. Doltra, M. Sitges, E. J. Vigmond, E. Zacur, V. Grau, B. Rodriguez, E. W. Remme, S. Niederer, P. Mortier, K. McLeod, M. Potse, E. Pueyo, A. Bueno-Orovio, and P. Lamata. The 'Digital Twin' to Enable the Vision of Precision Cardiology. *European Heart Journal*, Vol. 41, No. 48, 2020, pp. 4556–4564. https://doi.org/10.1093/EURHEARTJ/EHAA159.
10. Liu, Y., L. Zhang, Y. Yang, L. Zhou, L. Ren, F. Wang, R. Liu, Z. Pang, and M. J. Deen. A Novel Cloud-Based Framework for the Elderly Healthcare Services Using Digital Twin. *IEEE Access*, Vol. 7, 2019, pp. 49088–49101. https://doi.org/10.1109/ACCESS.2019.2909828.
11. Madni, A. M., C. C. Madni, and S. D. Lucero. Leveraging Digital Twin Technology in Model-Based Systems Engineering. *Systems 2019, Vol. 7, Page 7*, Vol. 7, No. 1, 2019, p. 7. https://doi.org/10.3390/SYSTEMS7010007.
12. Tao, F., F. Sui, A. Liu, Q. Qi, M. Zhang, B. Song, Z. Guo, S. C.-Y. Lu, and A. Y. C. Nee.

Digital Twin-Driven Product Design Framework. *https://doi.org/10.1080/00207543.2018.1443229*, Vol. 57, No. 12, 2018, pp. 3935–3953. https://doi.org/10.1080/00207543.2018.1443229.

13. Lidbe, A. D., E. G. Tedla, A. M. Hainen, and S. L. Jones. Feasibility Assessment for Implementing Adaptive Traffic Signal Control. *Journal of Transportation Engineering, Part A: Systems*, Vol. 145, No. 2, 2019, p. 05018002. https://doi.org/10.1061/JTEPBS.0000208.
14. Lidbe, A. D., E. G. Tedla, A. M. Hainen, A. Sullivan, and S. L. Jones. Comparative Assessment of Arterial Operations under Conventional Time-of-Day and Adaptive Traffic Signal Control. *Advances in Transportation Studies an international Journal Section A*, Vol. XLII, 2017, pp. 5–22.
15. Jayakrishnan, R., S. P. Mattingly, and M. G. McNally. Performance Study of SCOOT Traffic Control System with Non-Ideal Detectorization : Field Operational Test in the City of Anaheim. 2001.
16. Kergaye, C., A. Stevanovic, and P. T. Martin. An Evaluation of SCOOT and SCATS through Microsimulation. 2008.
17. Aljaafreh, A., and N. Al-Oudat. Optimized Timing Parameters for Real-Time Adaptive Traffic Signal Controller. *Proceedings - UKSim-AMSS 16th International Conference on Computer Modelling and Simulation, UKSim 2014*, 2014, pp. 244–247. https://doi.org/10.1109/UKSIM.2014.84.
18. Bhave, N., A. Dhagavkar, K. Dhande, M. Bana, and J. Joshi. Smart Signal - Adaptive Traffic Signal Control Using Reinforcement Learning and Object Detection. *Proceedings of the 3rd International Conference on I-SMAC IoT in Social, Mobile, Analytics and Cloud, I-SMAC 2019*, 2019, pp. 624–628. https://doi.org/10.1109/I-SMAC47947.2019.9032589.
19. McKenney, D., and T. White. Distributed and Adaptive Traffic Signal Control within a Realistic Traffic Simulation. *Engineering Applications of Artificial Intelligence*, Vol. 26, No. 1, 2013, pp. 574–583. https://doi.org/10.1016/J.ENGAPPAI.2012.04.008.
20. Gao, J., Y. Shen, J. Liu, M. Ito, and N. Shiratori. Adaptive Traffic Signal Control: Deep Reinforcement Learning Algorithm with Experience Replay and Target Network. 2017.
21. Zeng, J., J. Hu, and Y. Zhang. Adaptive Traffic Signal Control with Deep Recurrent Q-Learning. *IEEE Intelligent Vehicles Symposium, Proceedings*, Vol. 2018-June, 2018, pp. 1215–1220. https://doi.org/10.1109/IVS.2018.8500414.
22. Rachmadi, M. F., F. Al Afif, W. Jatmiko, P. Mursanto, E. A. Manggala, M. A. Ma'sum, and A. Wibowo. Adaptive Traffic Signal Control System Using Camera Sensor and Embedded System. *IEEE Region 10 Annual International Conference, Proceedings/TENCON*, 2011, pp. 1261–1265. https://doi.org/10.1109/TENCON.2011.6129009.
23. Aslani, M., M. S. Mesgari, and M. Wiering. Adaptive Traffic Signal Control with Actor-Critic Methods in a Real-World Traffic Network with Different Traffic Disruption Events. *Transportation Research Part C: Emerging Technologies*, Vol. 85, 2017, pp. 732–752. https://doi.org/10.1016/J.TRC.2017.09.020.
24. Abdulhai, B., R. Pringle, and G. J. Karakoulas. Reinforcement Learning for True Adaptive Traffic Signal Control. *Journal of Transportation Engineering*, Vol. 129, No. 3, 2003, pp. 278–285. https://doi.org/10.1061/(ASCE)0733-947X(2003)129:3(278).
25. Goodall, N. J., B. L. Smith, and B. (Brian) Park. Traffic Signal Control with Connected

Vehicles. *Transportation Research Record: Journal of the Transportation Research Board*, Vol. 2381, No. 1, 2013, pp. 65–72. https://doi.org/10.3141/2381-08.
26. Pandit, K., D. Ghosal, H. M. Zhang, and C. N. Chuah. Adaptive Traffic Signal Control with Vehicular Ad Hoc Networks. *IEEE Transactions on Vehicular Technology*, Vol. 62, No. 4, 2013, pp. 1459–1471. https://doi.org/10.1109/TVT.2013.2241460.
27. Talukder, M. A. S., A. D. Lidbe, E. G. Tedla, A. M. Hainen, and T. Atkison. Trajectory-Based Signal Control in Mixed Connected Vehicle Environments. *Journal of Transportation Engineering, Part A: Systems*, Vol. 147, No. 5, 2021, p. 04021016. https://doi.org/10.1061/JTEPBS.0000510.
28. Xiang, J., and Z. Chen. An Adaptive Traffic Signal Coordination Optimization Method Based on Vehicle-to-Infrastructure Communication. *Cluster Computing 2016 19:3*, Vol. 19, No. 3, 2016, pp. 1503–1514. https://doi.org/10.1007/S10586-016-0620-7.
29. Feng, Y., K. L. Head, S. Khoshmagham, and M. Zamanipour. A Real-Time Adaptive Signal Control in a Connected Vehicle Environment. *Transportation Research Part C: Emerging Technologies*, Vol. 55, 2015, pp. 460–473. https://doi.org/10.1016/j.trc.2015.01.007.
30. Kumar, S. A. P., R. Madhumathi, P. R. Chelliah, L. Tao, and S. Wang. A Novel Digital Twin-Centric Approach for Driver Intention Prediction and Traffic Congestion Avoidance. *Journal of Reliable Intelligent Environments 2018 4:4*, Vol. 4, No. 4, 2018, pp. 199–209. https://doi.org/10.1007/S40860-018-0069-Y.
31. Qadri, S. S. S. M., M. A. Gökçe, and E. Öner. State-of-Art Review of Traffic Signal Control Methods: Challenges and Opportunities. *European Transport Research Review 2020 12:1*, Vol. 12, No. 1, 2020, pp. 1–23. https://doi.org/10.1186/S12544-020-00439-1.
32. Wang, S., F. Zhang, and T. Qin. Research on the Construction of Highway Traffic Digital Twin System Based on 3D GIS Technology. *Journal of Physics: Conference Series*, Vol. 1802, No. 4, 2021, p. 042045. https://doi.org/10.1088/1742-6596/1802/4/042045.
33. Marai, O. El, T. Taleb, and J. Song. Roads Infrastructure Digital Twin: A Step Toward Smarter Cities Realization. *IEEE Network*, Vol. 35, No. 2, 2021, pp. 136–143. https://doi.org/10.1109/MNET.011.2000398.
34. Saroj, A. J. Development of a Real-Time Connected Corridor Data-Driven Digital Twin and Data Imputation Methods. 2020.
35. Hu, C., W. Fan, E. Zen, Z. Hang, F. Wang, L. Qi, and M. Z. A. Bhuiyan. A Digital Twin-Assisted Real-Time Traffic Data Prediction Method for 5G-Enabled Internet of Vehicles. *IEEE Transactions on Industrial Informatics*, 2021. https://doi.org/10.1109/TII.2021.3083596.
36. Dygalo, V., A. Keller, and A. Shcherbin. Principles of Application of Virtual and Physical Simulation Technology in Production of Digital Twin of Active Vehicle Safety Systems. *Transportation Research Procedia*, Vol. 50, 2020, pp. 121–129. https://doi.org/10.1016/J.TRPRO.2020.10.015.
37. Krajzewicz, D., 2010. Traffic simulation with SUMO–simulation of urban mobility. In Fundamentals of traffic simulation (pp. 269-293). Springer, New York, NY.
38. Traffic Control Interface (TraCI). Accessed on July 25, 2021. Available at: https://sumo.dlr.de/docs/TraCI.html
39. Manual, Highway Capacity. "HCM2010." Transportation Research Board, National Research Council, Washington, DC 1207 (2010).